\documentclass[prl,twocolumn]{revtex4}

\usepackage{psfrag}
\usepackage[T1]{fontenc}

\usepackage{times}
\usepackage{pifont}
\usepackage{ae,aecompl}
\usepackage{amsmath}
\usepackage{graphicx}

\begin{document}

\title{Fractal Structure of High-Temperature Graphs of O($N$) Models in
  Two Dimensions}
\author{Wolfhard Janke and Adriaan M. J. Schakel}
\affiliation{Institut f\"ur Theoretische Physik, Universit\"at Leipzig,
  Augustusplatz 10/11, 04109 Leipzig, Germany }

\begin{abstract}
  The critical behavior of the two-dimensional O($N)$ model close to
  criticality is shown to be encoded in the fractal structure of the
  high-temperature graphs of the model.  Based on Monte Carlo
  simulations and with the help of percolation theory, De Gennes'
  results for polymer rings, corresponding to the limit $N \to 0$, are
  generalized to random loops for arbitrary $-2 \leq N \leq 2$.  The
  loops are studied also close to their tricritical point, known as the
  $\Theta$ point in the context of polymers, where they collapse.  The
  corresponding fractal dimensions are argued to be in one-to-one
  correspondence with those at the critical point, leading to an analytic
  prediction for the magnetic scaling dimension at the O($N)$
  tricritical point.
\end{abstract}

\date{\today}

\maketitle

The high-temperature (HT) representation of the critical O($N$) spin
model naturally defines a loop gas, corresponding to a diagrammatic
expansion of the partition function in terms of closed graphs along the
bonds on the underlying lattice \cite{Stanley}.  In the limit $N\to 0$,
the loops reduce to closed self-avoiding random walks first considered
by de Gennes as a model for polymer rings in good solvents at
sufficiently high temperatures, so that the van der Waals attraction
between monomers is irrelevant \cite{deGennes}.  In his seminal paper,
de Gennes related the fractal structure of self-avoiding random walks to
the critical exponents of the O($N\to0$) model.  Invoking concepts from
percolation theory and recent Monte Carlo (MC) data
\cite{geoPotts,Dotsenko} of the HT representation of the two-dimensional
(2D) Ising model ($N=1$), we extend in this Letter de Gennes' result to
arbitrary $-2 \leq N \leq 2$ in 2D.  We consider the loops also close to
their tricritical point where they collapse, known as the $\Theta$ point
in the context of polymers.  We argue that the 2D fractal dimensions at
the tricritical point are in one-to-one correspondence with those at the
critical point, allowing us to also predict the magnetic scaling
dimension at the O($N)$ tricritical point.  We support our theoretical
prediction by comparing it with recent high-precision MC data
\cite{bloeteetal}.

A particularly simple representation of the O($N$) universality class is
specified by the partition function \cite{DMNS}
\begin{equation}
  \label{ZS}
  Z = \mathrm{Tr} \prod_{\langle \mathbf{x},\mathbf{x}' \rangle} (1 + K
  \mathbf{S}_\mathbf{x} \cdot \mathbf{S}_{\mathbf{x}'}),
\end{equation} 
where the product is over all nearest neighbor pairs, and the spins have
$N$ components $\mathbf{S}_\mathbf{x} = (S^1_\mathbf{x}, S^2_\mathbf{x},
\dots , S^N_\mathbf{x})$ and are of fixed length
$\mathbf{S}_\mathbf{x}^2=N$.  The trace Tr stands for the sum or
integral over all possible spin configurations.  The weighting factor is
obtained by truncating the more standard Boltzmann weight
$\exp(K\mathbf{S}_\mathbf{x} \cdot \mathbf{S}_{\mathbf{x}'})$.  This
choice mimics the weighting factor of the Ising model, where $\exp(\beta
S_\mathbf{x} S_{\mathbf{x}'}) \propto (1 + K S_\mathbf{x}
S_{\mathbf{x}'})$ with $K = \tanh \beta$.  When formulated on a
honeycomb lattice, which has coordination number $z=3$, the HT graphs of
the truncated model are automatically nonintersecting and self-avoiding. 
The partition function can then be written simply as a sum over all
possible closed graphs \cite{DMNS}, $Z = \sum_{\{G\}} K^b N^l$, with $b$
and $l$ the number of occupied bonds and separate loops forming the
graph $G$.  The parameter $K$ in the spin formulation (\ref{ZS}) appears
as bond fugacity in the loop model.  By mapping it onto a solid-on-solid
model, the critical exponents as well as the critical point were
determined exactly \cite{Nienhuis}.  

In the high-temperature phase, the HT graphs have a finite line tension
$\theta_\mathrm{G}$ and are exponentially suppressed.  A typical graph
configuration in this phase shows only a few small loops scattered
around the lattice.  Upon approaching the critical point from above, the
lattice starts to fill up with more and also larger graphs.  At the
critical point, the line tension vanishes, causing the exponential
suppression to disappear.  Graphs of all sizes now appear in the system
as they can grow without energy cost, i.e., the HT graphs proliferate. 
A graph spanning the lattice can be found irrespective of the lattice
size--much like the appearance of a spanning cluster at the percolation
threshold in percolation phenomena \cite{StauferAharony}.  The average
number density $\ell_b$ of graphs containing $b$ bonds takes
asymptotically a form similar to that of clusters in percolation theory,
\begin{equation}
  \label{ellb}
  \ell_b \sim b^{- \tau_\mathrm{G}} {\rm e}^{- \theta_\mathrm{G} b}, \quad 
  \theta_\mathrm{G} \propto (K-K_{\rm c})^{1/\sigma_\mathrm{G}},
\end{equation} 
with $\sigma_\mathrm{G}$ and $\tau_\mathrm{G}$ two exponents whose
values define the universality class.  The line tension vanishes upon
approaching the critical point at a pace determined by the exponent
$\sigma_\mathrm{G}$.  When present, this Boltzmann factor exponentially
suppresses large graphs.  The algebraic factor in the graph distribution
is an entropy factor, giving a measure of the number of ways a graph of
size $b$ can be embedded in the lattice.  The configurational entropy is
characterized by the exponent $\tau_\mathrm{G}$.  As in percolation
theory \cite{StauferAharony}, it is related to the fractal dimension
$D_\mathrm{G}$ of the HT graphs via
\begin{equation}
  \label{tau}
  \tau_\mathrm{G} = d/D_\mathrm{G} +1,
\end{equation} 
with $d=2$ the dimension of the lattice. 

When summed over all sizes, the graph distribution yields the scaling
part of the logarithm of the partition function,
\begin{equation}
  \label{lnZ} 
  \ln Z \sim \sum_b \ell_b. 
\end{equation} 
Each graph therefore contributes equally to the scaling part of the free
energy, irrespective of its size.

\begin{table}
  \begin{tabular}{l|rc|r|rcccc|ccc}
    \hline \hline & & & & & & & & \\[-.4cm]
    Model & $N$ & $\bar\kappa_-$ & $c$ & $\alpha$ & $\beta$ & $\gamma$ & $\eta$
    & $\nu$ & $D_{\rm G}$ & $\tau_\mathrm{G}$ & $\sigma_\mathrm{G}$ \\[.1cm]
    \hline & & & & & & & & & & \\[-.4cm]
    Gaussian & $-2$ & $\frac{1}{2}$ & $-2$ & $1$ & $0$ & $1$ & $0$ & $\frac{1}{2}$ &  $\frac{5}{4}$ & 
    $\frac{13}{5}$ &  $\frac{8}{5}$\\[.1cm]
    SAW & $0$ & $\frac{2}{3}$ & $0$ & $\frac{1}{2}$ & $\frac{5}{64}$ & 
    $\frac{43}{32}$ & $\frac{5}{24}$ & $\frac{3}{4}$  & $\frac{4}{3}$ &  $\frac{5}{2}$ &  $1$ \\[.1cm]
    Ising & $1$ & $\frac{3}{4}$ & $\frac{1}{2}$ 
    & $0$ & $\frac{1}{8}$ & $\frac{7}{4}$ &
    $\frac{1}{4}$ & $1$ &
    $\frac{11}{8}$ &  $\frac{27}{11}$ &  $\frac{8}{11}$ \\[.1cm]
    XY & $2$ & $1$ & $1$  & $-\infty$ & $\infty$ & $\infty$ & $\frac{1}{4}$ & $\infty$ 
    & $\frac{3}{2}$ &  $\frac{7}{3}$ &  $0$\\[.1cm] \hline \hline
  \end{tabular}
  \caption{Critical exponents of the two-dimensional critical O($N$) spin
    model, with $N=-2,0,1,2$, respectively, together with the parameter
    $\bar\kappa_-$ and the central charge $c= 1 - 6(1-\bar\kappa_-)^2/\bar\kappa_-$.  
    In addition, the fractal dimension
    $D_\mathrm{G}$ of the HT graphs  as well as the two exponents
    $\tau_\mathrm{G}, \sigma_\mathrm{G}$ characterizing their
    distribution are given.  For  $N=1$, the values of the latter two were recently 
    established numerically in Ref.~\cite{geoPotts}. 
    \label{table:On}}
\end{table}

Table \ref{table:On} summarizes the critical exponents and fractal
dimensions of the four most common O($N$) models.  The negative value
$N=-2$ corresponds to the noninteracting model, for which the critical
exponents take their Gaussian values \cite{BalianToulouse}.  In the
polymer limit $N\to 0$, first studied by de Gennes \cite{deGennes}, the
fractal dimension of the HT graphs is simply the inverse of the
correlation length exponent $\nu$.  In general, however, it follows from
Table \ref{table:On} that $D_\mathrm{G} \neq 1/\nu$.  To generalize de
Gennes' result, we note that in percolation theory a similar relation
between the fractal dimension $D$ of clusters and the correlation length
exponent involves the Fisher exponent $\sigma$, viz.\
\cite{StauferAharony}
\begin{equation}
  \label{nuD}
  D = 1/\sigma \nu. 
\end{equation} 
A closer look at de Gennes' derivation reveals that
$\sigma_\mathrm{G}=1$ in that case, implying that the result for
polymers in good solvents or self-avoiding random walks is consistent
with Eq.~(\ref{nuD}).  In a recent MC study of the HT graphs of
the 2D Ising model \cite{geoPotts}, we numerically found
the value $\sigma_\mathrm{G} = 0.732(6)$.  This estimate is within one
standard deviation from the value $\sigma_\mathrm{G} = 8/11 = 0.7273
\dots$ expected from Eq.~(\ref{nuD}), with $\nu=1$ and the fractal
dimension $D_\mathrm{G}=11/8$ appropriate for the Ising model. 

Parameterizing the 2D O($N$) models as \cite{Nienhuis_rev,SLE} $N = - 2
\cos(\pi/\bar\kappa_-)$, with $\frac{1}{2} \le \bar\kappa_- \le 1$, we
obtain from Eq.~(\ref{nuD}) $\sigma_\mathrm{G} = 8
(1-\bar\kappa_-)/(2+\bar\kappa_-)$, where use is made of the known
results \cite{Nienhuis} $1/\nu = 4 (1-\bar\kappa_-)$ and
\cite{Vanderzande}
\begin{equation}
  \label{dh}
  D_\mathrm{G} = 1+ \bar \kappa_-/2. 
\end{equation}   
The entropy exponent (\ref{tau}) follows similarly as $\tau_\mathrm{G} =
(6+ \bar\kappa_-)/(2+\bar\kappa_-)$, yielding $\tau_\mathrm{G}=5/2$ for
a self-avoiding random walk and $\tau_\mathrm{G}= 27/11 = 2.4546 \dots$
for the Ising model.  Through the exact enumeration and analysis of the
number of self-avoiding loops on a square lattice up to length 110, the
expected value $\tau_\mathrm{G}=5/2$ has been established numerically to
very high precision \cite{Jensen}.  In our MC study \cite{geoPotts}, we
numerically obtained the estimate $\tau_\mathrm{G} = 2.458(5)$ for the
Ising model in good agreement with the theoretical prediction.

The O($N$) spin-spin correlation function $G(\mathbf{x},\mathbf{x}') =
G(\mathbf{x}-\mathbf{x}')$ is represented diagrammatically by a modified
partition function, obtained by requiring that the two sites
$\mathbf{x}$ and $\mathbf{x}'$ are connected by an open HT graph
$\Gamma$ \cite{Stanley}.  On a honeycomb lattice, the scaling part of
the correlation function is given by the connected graphs
\begin{equation}
\label{GR}
  G(\mathbf{x},\mathbf{x}') \sim \sum_{\{\Gamma\}} K^b = \sum_b
  z_b(\mathbf{x},\mathbf{x}') K^b,
\end{equation} 
where $z_b(\mathbf{x},\mathbf{x}')$ is the number of (open)
nonintersecting and self-avoiding graphs along $b$ bonds connecting
$\mathbf{x}$ and $\mathbf{x}'$.  It is related to the graph distribution
(\ref{ellb}) through $ \ell_b = (1/V b) \sum_\mathbf{x}
z_b(\mathbf{x},\mathbf{x})$, with $V$ the lattice volume.  Since
$z_b(\mathbf{x},\mathbf{x})$ refers to closed graphs starting and ending
at $\mathbf{x}$, the factor $1/b$ is included to prevent overcounting as
a given loop can be traced out starting at any lattice point along that
loop.  Strictly speaking, $G(\mathbf{x},\mathbf{x}') < \sum_{\{\Gamma\}}
K^b$ as the cancellation of the disconnected graphs in the numerator and
$Z$ in the denominator, required for an equality, is not complete: For a
given open graph $\Gamma$, certain loops present in $Z$ are forbidden in
the modified partition function as they would intersect $\Gamma$, or
occupy bonds belonging to it, which is not allowed.  In other words, the
presence of an open graph influences the loop gas and \textit{vice
versa}.  Since each loop carries a factor $N$, the loop gas is absent in
the limit $N \to 0$.  The inequality in Eq.~(\ref{GR}) then becomes an
equality and the open graphs become ordinary self-avoiding random walks
with $D_\mathrm{G}=4/3$.  For $N>0$, the loops obstruct the formation of
graphs connecting the two endpoints, so that the fractal dimension of
these self-avoiding graphs on the honeycomb lattice is larger.

The magnetic susceptibility $\chi$ follows as $\chi = \sum_{\mathbf{x}'}
G(\mathbf{x},\mathbf{x}') \sim \sum_b z_b K^b$, with $z_b =
\sum_{\mathbf{x}'} z_b(\mathbf{x},\mathbf{x}')$ the number of open
graphs of size $b$ starting at $\mathbf{x}$ and ending at an arbitrary
lattice point.  For the susceptibility to diverge with the correct
exponent $\gamma$, it must behave close to the critical point as
\begin{equation}
  \label{asymp} 
  \chi \sim  \sum_b b^{\sigma_\mathrm{G} \gamma-1}
  \mathrm{e}^{-\theta_\mathrm{G} b},
\end{equation} 
for large $b$, where like in the closed graph distribution (\ref{ellb}), the
Boltzmann factor suppresses large graphs as long as the line tension
$\theta_\mathrm{G}$ is finite.  Indeed, replacing the summation over the
HT graph size $b$ with an integration, we find $\chi \sim
|K-K_\mathrm{c}|^{-\gamma}$.  The asymptotic form (\ref{asymp})
generalizes the de Gennes result for $N\to 0$ with $\sigma_\mathrm{G}=1$
to arbitrary $-2 \leq N \leq 2$ with $\sigma_ \mathrm{G} \neq 1$.  Note
that on account of Fisher's scaling relation $\gamma = (2-\eta) \nu$,
the combination $\sigma_\mathrm{G} \gamma$ in Eq.~(\ref{asymp})
satisfies $\sigma_\mathrm{G} \gamma = (2-\eta)/D_\mathrm{G}$.

The ratio of $z_b(\mathbf{x},\mathbf{x}')$ and $z_b$ defines the
probability $P_b(\mathbf{x},\mathbf{x}')$ of finding a graph connecting
$\mathbf{x}$ and $\mathbf{x}'$ along $b$ bonds \cite{Cloizeaux}.  On
general grounds, it scales at criticality as ($d=2$):
\begin{equation}
  \label{P}
  P_b(\mathbf{x},\mathbf{x}') = z_b(\mathbf{x},\mathbf{x}')/z_b \sim
  b^{-d/D_\mathrm{G}} \, \mathsf{P} \left(
    |\mathbf{x}-\mathbf{x}'|/b^{1/D_\mathrm{G}} \right),
\end{equation} 
with $\mathsf{P}$ a scaling function.  Since \textit{at} the critical
point, $z_b K^b_\mathrm{c} \sim b^{\sigma_\mathrm{G} \gamma-1}$
according to Eq.~(\ref{asymp}), we obtain for the correlation function
\begin{equation}
  \label{G}
  G(\mathbf{x},\mathbf{x}') \sim  \sum_b z_b
  P_b(\mathbf{x},\mathbf{x}') K_\mathrm{c}^b \sim 
  1/|\mathbf{x}-\mathbf{x}'|^{d-2 + \eta},
\end{equation} 
where use is made of Eq.~(\ref{nuD}) and Fisher's scaling relation.
Equation (\ref{G}) is the standard definition of the critical exponent
$\eta$, whose exact value is given by \cite{Nienhuis}
\begin{equation}
  \label{etakappa}
  \eta=2-3/4\bar\kappa_- - \bar\kappa_-,
\end{equation}  
and thus provides a consistency check.  Also, with $\nu =
(\tau_\mathrm{G}-1)/d \sigma_\mathrm{G}$, as is implied by
Eqs.~(\ref{nuD}) and (\ref{tau}), Eq.~(\ref{lnZ}) yields the scaling
relation $d \nu = 2 - \alpha$, where $\alpha$ determines the scaling
behavior of the free energy close to the critical point, $\ln Z \sim
|K-K_\mathrm{c}|^{2-\alpha}$.  Apart from $\nu$ and $\alpha$, the
exponents have no simple dependence on the graph distribution exponents
$\sigma_\mathrm{G}, \tau_\mathrm{G}$.  This is because the operator
whose scaling dimension $y_\mathrm{G}$ is given by the fractal dimension
$D_\mathrm{G}$ of the HT graphs is not a simple one, consisting of two
spin components at the same site $S_\mathbf{x}^i S_\mathbf{x}^j$ which
measures the tendency of spins to align \cite{Nienhuis}.

In our MC study \cite{geoPotts}, rather than determining the
scaling dimension $y_\mathrm{G}$ directly, we measured the so-called
percolation strength $P_\infty$, giving the fraction of bonds in the
largest graph.  This observable obeys the finite-size scaling relation
$P_\infty \sim L^{-\beta_{\rm G}/\nu}$, where the exponent $\beta_{\rm
  G}$ is related to $y_\mathrm{G}$ via $y_\mathrm{G} = d -
\beta_\mathrm{G}/\nu$.  We found $\beta_{\rm G} = 0.626(7)$ for the
Ising model in perfect agreement with the value $\beta_{\rm
  G}=5/8=0.625$, leading to $y_\mathrm{G} = 11/8$, which coincides with
the fractal dimension $D_\mathrm{G}$ of the HT graphs. 

We next extend our results for the HT graphs at the \textit{critical}
point to the \textit{tricritical} point.  It is generally accepted that
including vacancies in the O($N$) model gives in addition to critical
behavior rise to also tricritical behavior.  By gradually increasing the
activity of the vacancies, the continuous O($N$) phase transition is
eventually driven first order.  The endpoint where this happens is a
tricritical point.  In the context of polymers ($N \to 0$), the latter
obtains by lowering the temperature to the $\Theta$ point where the
increasingly important van der Waals attraction between monomers causes
the polymer chain to collapse.  Coniglio \textit{et
  al.\/}~\cite{Coniglioetal} argued that a polymer ring at the $\Theta$
tricritical point is equivalent to the hull of a percolation cluster.
With the known dimension for the percolation hull \cite{SDFK}, the
analogy then gives $D_\mathrm{G}^\mathrm{t}=\frac{7}{4}$ as fractal
dimension for a polymer chain at the $\Theta$ point (the superscript
``t'' refers to the tricritical point).  It also implies that polymer
chains at the critical and tricritical point share the same central
charge $c$ because both the O($N \to 0$) model and percolation have
$c=0$ \cite{DStheta}.

To generalize the analogy found by Coniglio \textit{et
  al.\/}~\cite{Coniglioetal}, we consider the $Q$-state Potts model,
which in the limit $Q\to 1$ describes ordinary, uncorrelated percolation
\cite{Potts}.  In the Fortuin-Kasteleyn (FK) formulation \cite{FK}, the
model with $Q>1$ is mapped onto a \textit{correlated} percolation
problem.  Clusters are formed by lumping together with a certain
temperature-dependent probability nearest neighbor spins in the same
spin state.  These so-called FK clusters percolate at the critical
temperature and their fractal structure encodes the entire critical
behavior.  In 2D (and in 2D only), also the geometrical clusters, formed
by unconditionally lumping together nearest neighbor spins in the same
spin state, percolate at the critical temperature.  Their fractal
structure encodes the \textit{tricritical} $Q_\mathrm{t}$-state Potts
behavior, which emerges when enlarging the pure model to include
vacancies.  The tricritical point shares the same central charge $c$
with the critical point, but apart from $c=1$, $Q(c) \neq
Q_\mathrm{t}(c)$.  Both fractal structures and thus both critical
behaviors are intimately related, being connected by replacing
$\bar\kappa_-$ with $\bar\kappa_+ = 1/\bar\kappa_- \geq 1$ in the
appropriate expressions (for details, see Ref.~[\onlinecite{geoPotts}]
or Ref.~[\onlinecite{DBN}], where similar conclusions were reached
independently). This map conserves the central charge. With increasing
$Q$ or $c$, the critical and tricritical points approach each other
until merging at $c=1$, where $Q=Q_\mathrm{t}=4$ and the scaling
behaviors of FK and geometrical clusters coincide.  Beyond $Q=4$, the
transition becomes discontinuous. 

The hulls of the geometrical clusters of the $Q(c)$-state Potts model at
the same time represent the HT graphs of the critical O($N$) model, with
$0\leq N(c) \leq 2$, sharing the same central charge
\cite{DS89,geoPotts}.  Since the geometrical clusters encode the
tricritical Potts behavior, while the FK clusters encode the critical
Potts behavior (characterized by the same central charge), it is natural
to expect the FK hulls to represent the HT loop gas not at the critical,
but at the tricritical point.  For the special case of polymers ($N \to
0$), this reproduces the result by Coniglio \textit{et
al.\/}~\cite{Coniglioetal}.  Note that in both Potts and O($N$) models,
including vacancies leads to tricritical behavior.  However, the
critical and tricritical points get interchanged when passing from one
model to the other.  The fractal dimension $D_\mathrm{G}^\mathrm{t}$ of
the tricritical loops follows by applying the central-charge conserving
map $\bar\kappa_- \to \bar\kappa_+ = 1/\bar\kappa_-$ to Eq.~(\ref{dh}).
We submit that the exponent $\eta_\mathrm{t}$ follows in the same way
from Eq.~(\ref{etakappa}). Given $\eta_\mathrm{t}$, the scaling
relations then yield values for the ratios
$\beta_\mathrm{t}/\nu_\mathrm{t}$ and
$\gamma_\mathrm{t}/\nu_\mathrm{t}$.

\begin{figure}
\centering
\includegraphics[width=0.45\textwidth]{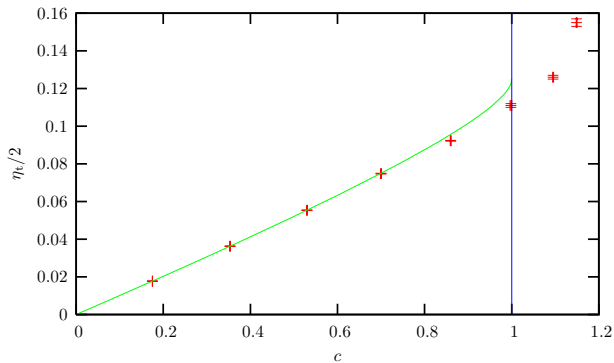}
\caption{Our analytic prediction for the tricritical exponent
  $\eta_\mathrm{t}$ (solid line) as a function of the central charge $c$
  compared with high-precision MC data (marks with error bars)
  \cite{bloeteetal}. Our prediction extends only to $c=1$, indicated by
  the vertical line, while the numerical results continue to the
  tricritical O(2) model with $c=1.149(1)$ \cite{bloeteetal}.
  \label{fig:xh_c}}
\end{figure}

To verify these predictions we rewrite our results, given as a function
of $\bar \kappa_-$, as entries in the Kac table
\begin{equation}
  h_{p,q} = \frac{[(m+1)p -m q]^2 -1}{4m(m+1)}. 
\end{equation} 
Here, the parameter $m$ is related to $\bar\kappa_-$ and the central
charge via $\bar\kappa_- = m/(1+m)$ and $c = 1 - 6/m(m+1)$,
respectively, while the central-charge conserving map becomes $m \to
-m-1$.  Specifically, for $0 \leq c \leq 1$
\begin{equation}
  \label{Dt}
  D_\mathrm{G}^\mathrm{t} =  1+ 1/2\bar \kappa_- = 2 -2 h_{m,m},
\end{equation} 
leading to the correct result \cite{DStheta}
$D_\mathrm{G}^\mathrm{t}=2-2h_{2,2}=\frac{7}{4}$ for a polymer at the
$\Theta$ point ($m=2$), and
\begin{equation}
\label{etat}
  \eta_\mathrm{t} = 2-3\bar\kappa_-/4 - 1/\bar\kappa_- = 4 h_{m/2,m/2}. 
\end{equation} 
The right hand of this equation is in agreement with the results
$\eta_\mathrm{t}=4 h_{1,1}=0$ \cite{DStheta} for the $\Theta$ point
($m=2$) and $\eta_\mathrm{t}=4 h_{2,2} = \frac{3}{20}$
\cite{Zamolodchikov} for the tricritical Ising model ($m=4$).  After
circulating a draft of this paper, we have been informed about a recent
high-precision MC study of the tricritical O($N$) model in
Ref.~\cite{bloeteetal} previously unavailable to us, in which the
authors extend earlier numerical work on the tricritical
O($\frac{1}{2}$) model \cite{GBN}.  In Fig.~\ref{fig:xh_c}, we compare
our theoretical prediction (\ref{etat}) for $\eta_\mathrm{t}$ with the
magnetic scaling dimension
$x^\mathrm{t}_h=2-y^\mathrm{t}_h=\eta_\mathrm{t}/2$ obtained numerically
in that study.  For $0 \leq c \leq \frac{7}{10}$, the MC data
are within one standard deviation of our prediction.  Beyond the
tricritical Ising model ($c=\frac{7}{10}$) the numerical data start
deviating from our analytic result.  A detailed future investigation is
required to clarify this discrepancy.

The tricritical HT graphs, representing simultaneously the hulls of FK
clusters, have a distribution again of the form (\ref{ellb}),
characterized by two exponents
$\sigma_\mathrm{G}^\mathrm{t}, \tau_\mathrm{G}^\mathrm{t}$.  Given our
result (\ref{Dt}) for the fractal dimension of these graphs,
$\tau_\mathrm{G}^\mathrm{t}$ is determined exactly through
Eq.~(\ref{tau}) with $\tau_\mathrm{G}$ and $D_\mathrm{G}$ replaced by
their tricritical counterparts $\tau^\mathrm{t}_\mathrm{G}$ and
$D^\mathrm{t}_\mathrm{G}$.

In conclusion, we have shown that for $-2 \leq N \leq 2$ the fractal
structure of 2D HT graphs of the O($N$) spin model encodes the O($N$)
critical behavior.  We thereby extended de Gennes' result for
self-avoiding loops in the limit $N \to 0$ to random loops for arbitrary
$-2 \leq N \leq 2$.  We studied the loops also close to the point where
they collapse, corresponding to the HT representation of the tricritical
O($N$) model.  The fractal structure of the tricritical loops was argued
to be in one-to-one correspondence with that of the critical loops,
allowing us to also predict the magnetic scaling dimension
$x^\mathrm{t}_h =\eta_\mathrm{t}/2 = 2 h_{m/2,m/2}$ at the O($N)$
tricritical point, in very good agreement with recent MC data in the
range $0 \leq c \lesssim 0.7$.

%
The authors thank H. Bl\"ote, C. von Ferber, Y. Holovatch, and B. 
Nienhuis for useful discussions and communications.  This work is
partially supported by the DFG grant No. JA 483/17-3 and by the
German-Israel Foundation (GIF) under grant No.\ I-653-181.14/1999.

\end{document}